\documentclass[3p,times]{elsarticle}

\usepackage{amssymb}

\usepackage[figuresright]{rotating}

\begin{document}

\begin{frontmatter}




\title{Valence Band Electronic Structure of Nb$_2$Pd$_{1.2}$Se$_5$ and 
Nb$_2$Pd$_{0.95}$S$_5$ Superconductors}


\author[one,two]{H. Lohani}
\author[one]{ P. Mishra}
\author[three]{R. Goyal}
\author[three]{V. P. S. Awana}
\author[one,two]{B. R. Sekhar\corref{correspondingauthor}\fnref{label2}}
\cortext[correspondingauthor]{Corresponding author}
\ead{sekhar@iopb.res.in}
\fntext[label2]{ Institute of Physics, Sachivalaya Marg,  Bhubaneswar-751 005, India\\      
                         Phone:  91-674-2306412 (Office), Fax:    91-674-2300142
}
\address[one]{Institute of Physics, Sachivalaya Marg, Bhubaneswar 751005,
India.}
\address[two]{Homi Bhabha National Institute, Training School Complex, Anushakti
Nagar, Mumbai 400085, India}
\address[three]{National Physical Laboratory(CSIR),
Dr. K. S. Krishnan Road, New Delhi 110012, India.}
\begin{abstract}
We present a comparative study of our valence band photoemission results 
on Nb$_2$Pd$_{1.2}$Se$_5$ and Nb$_2$Pd$_{0.95}$S$_5$ superconductors which 
is supported by our DFT based electronic structure calculations.
We observe that the VB spectra of both the compounds are qualitatively similar, except slight
difference in the binding energy position of  all  features between the two compounds which could be the result of different
electronegativity of Se and S atom. The calculated density of states(DOS) reveal that the VB features are mainly composed 
of Pd-Se/S hybridized states. The nature of DOS originating from the distinctly coordinated Pd atoms
is different. Further, the involvement of the various Pd-4d and Nb-4d states in crossing of
Fermi level(E$_f$) signifies the multiband character of these compounds. In addition,
we find a temperature dependent pseudogap in Nb$_2$Pd$_{0.95}$S$_5$ which is absent in
Nb$_2$Pd$_{1.2}$Se$_5$.
\end{abstract}


\end{frontmatter}


\section{Introduction}
\label{1}
Superconducting state of the matter remains as an exciting as well as 
extensively studied topic from the past. Recently discovered superconductors 
(SC), like Fe-pnictides\cite{stew}, Fe-chalcogenides\cite{taka}, 
MgB$_2$\cite{patn}, SrRuO$_4$\cite{maeno}, organic SC\cite{norman,olsen} have 
added new concepts, like multiband effects, admixture of spin singlet and 
triplet pairing, spin orbit coupling(SOC) effects {\it etc}. to the understanding 
of the superconducting phenomena. Among them Pd based ternary chalcogenides, 
like Nb$_2$Pd$_{0.95}$S$_5$\cite{Jha,Zhang}, Nb$_2$PdSe$_5$\cite{khim}, 
Ta$_2$PdS$_5$\cite{Lu}, Ta$_2$Pd$_{0.97}$S$_6$\cite{Tiwari}, 
Ta$_2$Pd$_{0.97}$Te$_6$\cite{goyal} and Ta$_4$Pd$_3$Te$_{16}$\cite{jiao} 
show some peculiar behaviors arising from their low dimensional structure and 
strong SOC.

Nb$_2$Pd$_{1.2}$Se$_5$ and Nb$_2$Pd$_{0.95}$S$_5$ are isomorphic and have 
their superconducting critical temperature, T$_{c}$ $\sim$ 6 K\cite{Jha,khim}. Although, 
Nb$_2$Pd$_{0.95}$S$_5$ shows a Fermi liquid behavior at low temperatures, the 
value of Sommerfiled coefficient estimated for Nb$_2$Pd$_{1.2}$Se$_5$ and 
Nb$_2$Pd$_{0.95}$S$_5$, 15.7 and 32 mJ/mol-K$^2$ respectively, indicate a moderately and 
strongly coupled electronic interaction respectively in them. Heat capacity 
measurements have shown signatures of multiband superconducting behavior in 
both the compounds which has been well described by the two band-$\alpha$ 
model.  The calculated Fermi surface(FS) of both the 
compounds exhibit sheets of electron and hole character of different 
dimensions. Nesting between these 1-D like sheets of FS thought to give rise  
various density wave orderings, like charge density wave (CDW) and spin 
density wave (SDW) in these systems\cite{David,Zhang}. These interesting behaviour
could be originated from  their complex atomic packing as hybridization strength
among different atomic orbitals mainly depends on structural geometry.

Theoretical investigations of the electronic structure of some of these 
superconducting transition metal chalcogenides, like Nb$_2$PdSe$_5$\cite{khim}, 
Nb$_2$PdS$_5$\cite{Zhang}, Ta$_2$PdS$_5$\cite{Singh}, 
Ta$_4$Pd$_3$Te$_{16}$\cite{David} confirm the multiband nature of these 
compounds. However, till date no direct experimental studies like photoemission
 has been reported on the valence band (VB) electronic structure of these class of SCs. 
In this paper, we present a comparative study of our angle integrated ultra violet photoemisson results 
supported by our DFT based calculations on Nb$_2$Pd$_{1.2}$Se$_5$ and 
Nb$_2$Pd$_{0.95}$S$_5$.  Our measurements show the VB spectra of both the compounds are qualitatively similar
with slight difference in binding energy position of  all the features  between the two compounds which  could be 
originated due to different electronegativity of Se and S atom. The calculated density of states(DOS) describe the
VB features are mainly composed of Pd-Se/S hybridized states. The different nature of DOS originating from the differently 
coordinated atoms signifies an important role of the complex structural geometry and multiband effects in the electronic
structure of these compounds.

\section{Experimental and calculation details}
\label{2}
Polycrystalline samples of Nb$_2$Pd$_{1.2}$Se$_5$ and Nb$_2$Pd$_{0.95}$S$_5$ 
were synthesized via solid state reaction route. Structural and physical 
properties were studied and reported earlier\cite{Jha,khim}. Photoelectron 
spectroscopic studies were performed in angle integrated mode using a 
hemispherical SCIENTA-R3000 analyzer and a monochromatized He source 
(SCIENTA-VUV5040). The photon flux was of the order of 10$^{16}$ 
photons/s/steradian with a beam spot of 2 mm in diameter. Fermi energy (E$_f$) 
for all the spectra were calibrated by measuring the E$_f$ of a freshly 
evaporated Ag film onto the sample holder. The total energy resolution, 
estimated from the width of the Fermi edge, was about 27 meV for the He I 
excitation. All the photoemission measurements were performed inside the 
analysis chamber under a base vacuum of $\sim$ $3.0$ $\times$ $10^{-10}$ mbar. 
The polycrystalline samples were repeatedly scraped using a diamond file 
inside the preparation chamber with a base vacuum of $\sim$ $5.0$ 
$\times$ $10^{-10}$ mbar and the spectra were taken within $1$ hour, so as to 
avoid any surface degradation. All measurements were repeated many times to 
ensure the reproducibility of the spectra. For the temperature dependent 
measurements, the samples were cooled by pumping liquid nitrogen through the 
sample manipulator fitted with a cryostat. Sample temperatures were measured 
using a silicon diode sensor touching the bottom of the stainless steel 
sample plate. The low temperature photoemission measurements at $77$ K were 
performed immediately after the cleaning the sample surfaces followed by the 
room temperature measurements.

First-principles calculations were performed by using plain wave basis set 
inherent in Quantum Espresso (QE)\cite{qe}. Many electron exchange-correlation 
energy approximated by Perdew-Burke-Ernzerhof (PBE) functional
\cite{Ernzerhof,Wang,Chevary} in a scalar relativistic, ultrasoft 
pseudopotential was employed\cite{Vanderbilt}. Fine mesh of k-points with 
Gaussian smearing of the order of 0.0001 Ry was used for sampling the 
Brillouin zone integration. Kinetic energy and charge density cut-off were set 
to 180 Ry and 900 Ry respectively. Experimental lattice parameters and atomic 
coordinates of Nb$_2$Pd$_{1.2}$Se$_5$\cite{khim} and 
Nb$_2$Pd$_{0.95}$S$_5$\cite{Jha} after relaxation under damped (Beeman) 
dynamics with respect to both ionic coordinates and the lattice vector, were 
used in the calculations. 

\section{Results and discussion}
\label{3}
\begin{figure}
\includegraphics[width=7cm,keepaspectratio]{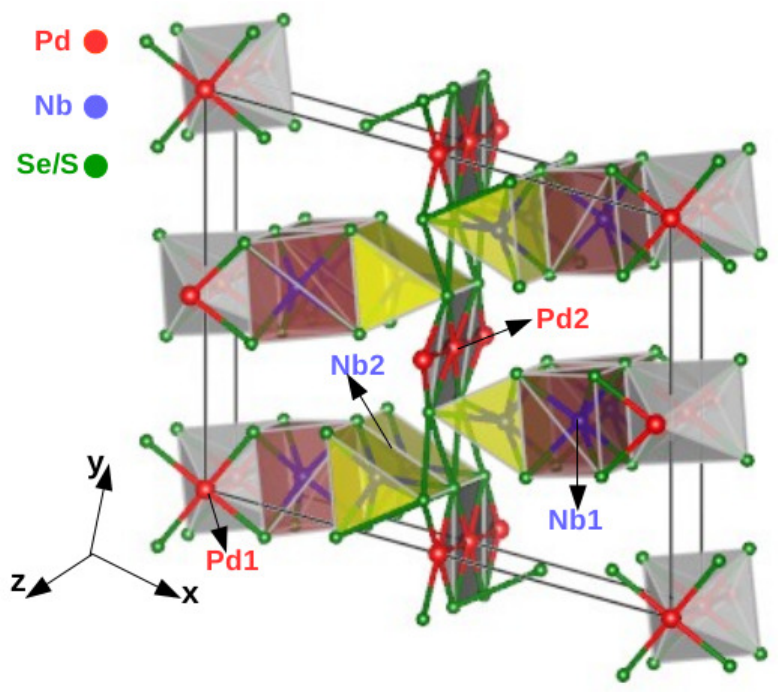}
\caption{\label{crystal}(a) Crystal structure of Nb$_2$Pd(Se/S)$_5$,
 Pd and Nb atoms at two inequivalent sites  are marked.}
\end{figure}

Fig.\ref{crystal} shows crystal structure of Nb$_2$Pd(Se/S)$_5$ which 
crystallizes in centrosymmetric structure with space group symmetry
C$_{2/m}$(\# 12). The lattice parameters of Nb$_2$Pd(Se/S)$_5$ are
a = 12.788/12.134 \AA{}, b = 3.406/3.277 \AA{}, c = 15.567/15.023 \AA{}
and $\beta$ = 101.63$^{\circ}$/103.23$^{\circ}$. The further details of
crystal structures regarding Wyckoff poisitions, site symmetry and fractional
occupancies can be obtained from Ref.\cite{khim}/\cite{Jha} for Nb$_2$Pd(Se/S)$_5$.
Both the structures mainly comprise of chains of Pd and Nb polyhedra with Se/S 
atoms. Pd and Nb atoms are placed at two inequivalent sites with different 
coordination environment denoted by Pd1, Pd2 and Nb1, Nb2. The atoms denoted by Pd1 
are arranged in columns facing their square planarly atoms and in octahedral 
coordination, whereas, the Pd2 site has a distorted rhombohedral prismatic 
environment\cite{douglas}. The Nb atoms form edge sharing and face sharing 
columns at Nb1 and Nb2 sites respectively and exist in a trigonal prismatic
 coordination. In packing these Pd and Nb centered polyhedra several octahedral and 
tetrahedral vacant sites are created. These vacancies lead to site dependent
 variations in the interorbital hybrdization of transition metal atoms with their nearest 
neighbour (nn) chalcogen atoms due to their different coordination and bond 
lengths which consequently reflect on the electronic structure of both the 
compounds.

\begin{figure}
\includegraphics[width=7cm,keepaspectratio]{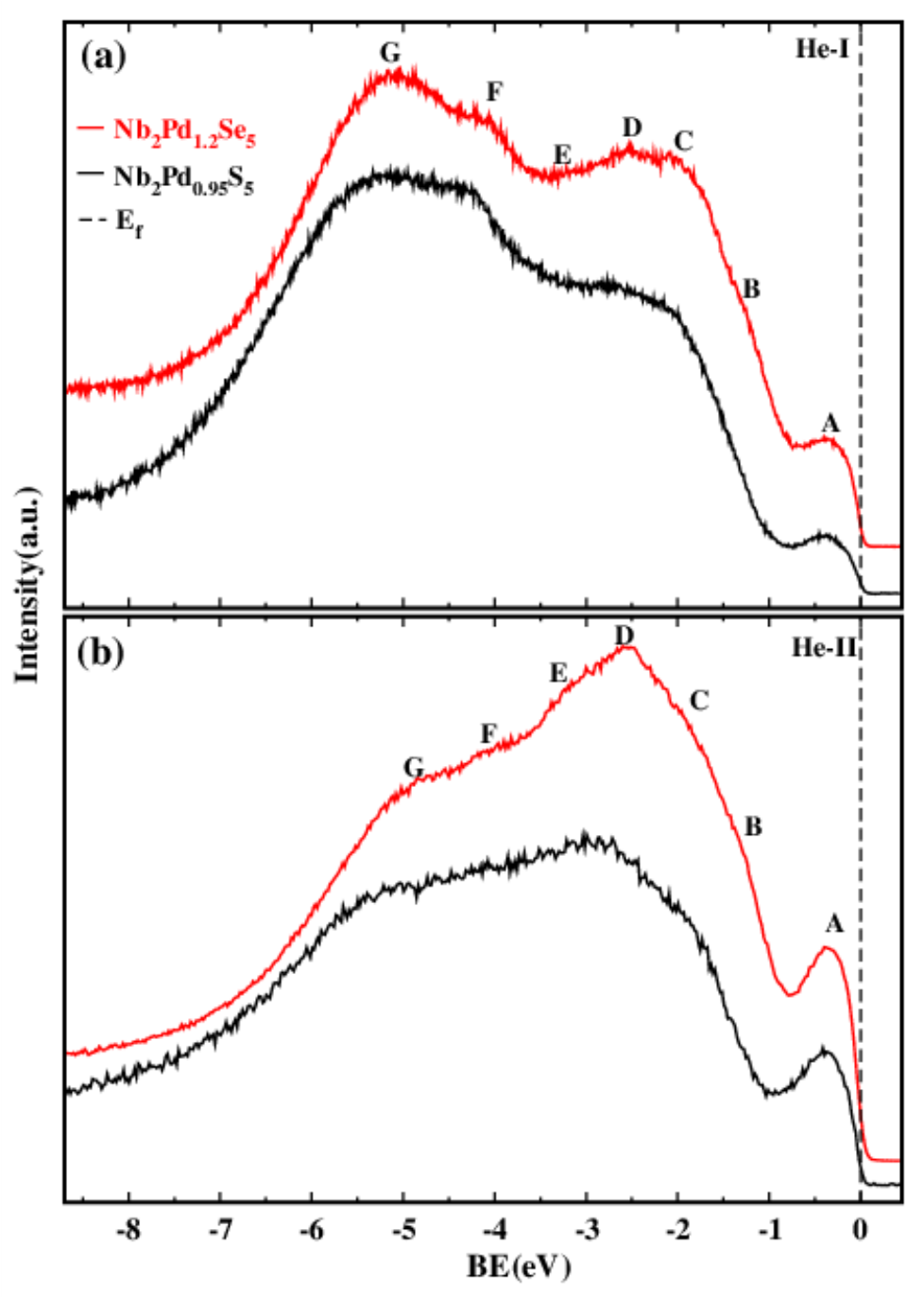}
\caption{\label{valcom}(a) and (b)  valence band spectra of 
Nb$_2$Pd$_{1.2}$Se$_5$(Red) and Nb$_2$Pd$_{0.95}$S$_5$(Black) obtained by 
using HeI and HeII excitation energy at 300 K respectively. }
\end{figure}

Fig.\ref{valcom}(a) shows valence band (VB) spectra of Nb$_2$Pd$_{1.2}$Se$_5$ 
and Nb$_2$Pd$_{0.95}$S$_5$ taken at HeI (21.2 eV) photon energy. In 
Nb$_2$Pd$_{1.2}$Se$_5$, seven features are observed at binding energy (BE) E$_b$ = -0.27, 
-1.25, -1.99, -2.53, -3.27, -4.03 and -5.08 eV which are marked by A to G 
respectively. The features D, F and G are more intense compared to E and C 
while the near E$_f$ feature A appears as a hump type structure. The VB 
spectra of Nb$_2$Pd$_{0.95}$S$_5$ is similar to that of 
Nb$_2$Pd$_{1.2}$Se$_5$, except that all the features occur at slightly higher 
BE probably due to its higher resistivity which could be
originating from the large electronegativity of S anion in comparison to Se 
anion. In Fig.\ref{valcom}(b) the VB spectra of the two compounds taken at He 
II (40.8 eV) excitation energy is presented. The He II spectra show sharp 
enhancement for features A, D and E due to change in the photoemission matrix 
elements with variation in photon energy. Atomic photoionization cross section 
($\sigma$) of Pd-4d relative to Se-4p (S-3p) is $\sim$ 8 (14) and 146 (134) 
for He I and He II respectively\cite{lindau}. This indicates that features A, 
D and E are mainly dominated by Pd-4d states in both these compounds. In the 
cases of both He I and He II, the near E$_f$ spectral weight is lesser in 
Nb$_2$Pd$_{0.95}$S$_5$ compared to Nb$_2$Pd$_{1.2}$Se$_5$. This result is 
again consistent with their conductivity nature which is smaller in case of 
Nb$_2$Pd$_{0.95}$S$_5$\cite{Jha} in comparison to Nb$_2$Pd$_{1.2}$Se$_5$\cite{khim}. 

\begin{figure}
\includegraphics[width=7cm,keepaspectratio]{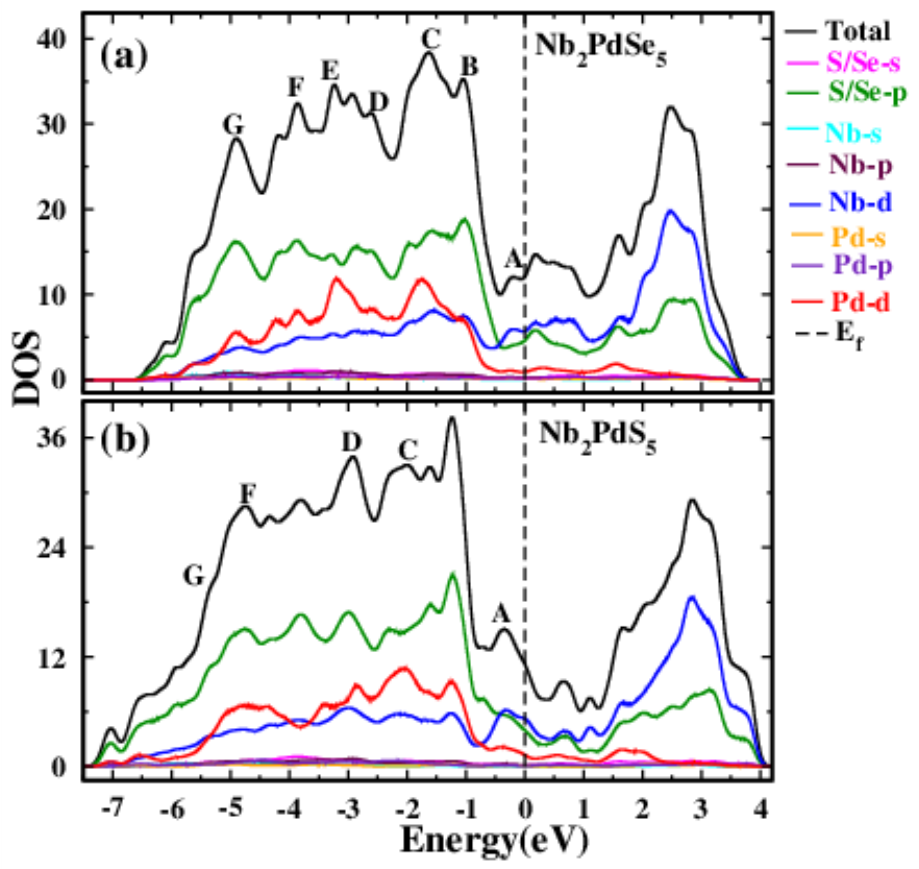}
\caption{\label{dos}(a) and (b)  calculated Total DOS with different 
atomic contributions of Nb$_2$PdSe$_5$ and Nb$_2$PdS$_5$ 
respectively.}
\end{figure}

First principles based calculations have been performed on systems with 
comparable compositions(Nb$_2$PdSe$_5$, and Nb$_2$PdS$_5$) in order to have a guidance in interpreting the 
experimental results. Fig.\ref{dos}(a) and (b) show calculated DOS along with 
different atomic contributions of Nb$_2$PdSe$_5$ and Nb$_2$PdS$_5$ 
respectively. The strongly hybridized states of Pd-4d with their nn Se(S)-p are 
dominant in  VB part $\sim$ BE range E$_b$ = -7.0 to -1.0 eV, whereas sufficient states originating
from Nb-4d and Se(S)-p hybridization are present in the near E$_f$ region of DOS. These Nb-4d and Se(S)-p
hybridized states are also major component in  conduction band(CB) region of the DOS in Nb$_2$PdSe$_5$(Nb$_2$PdS$_5$).
The lower BE of calculated peaks in Nb$_2$PdSe$_5$, in comparison to those of 
Nb$_2$PdS$_5$, are  in agreement with the trends observed in the 
experimental data of the VB spectra. Similarly, the DOS of both the compounds consist 
various  peak type structures which are not well resolved in the experimental VB spectra.
Possible reason for this could be inelastic scattering of the photoelectrons. 
However, the calculated  peaks, which are close to the experimentally observed VB features
are labeled with same notation as used in the experimental VB spectra(Fig.\ref{valcom}). 
These results describe the features E, and D mainly derived from Pd-4d states which is
consistent with the significant enhancement of these features  due to
variation in  $\sigma$ of Pd-4d in He II spectra.

\begin{figure}
\includegraphics[width=10cm,keepaspectratio]{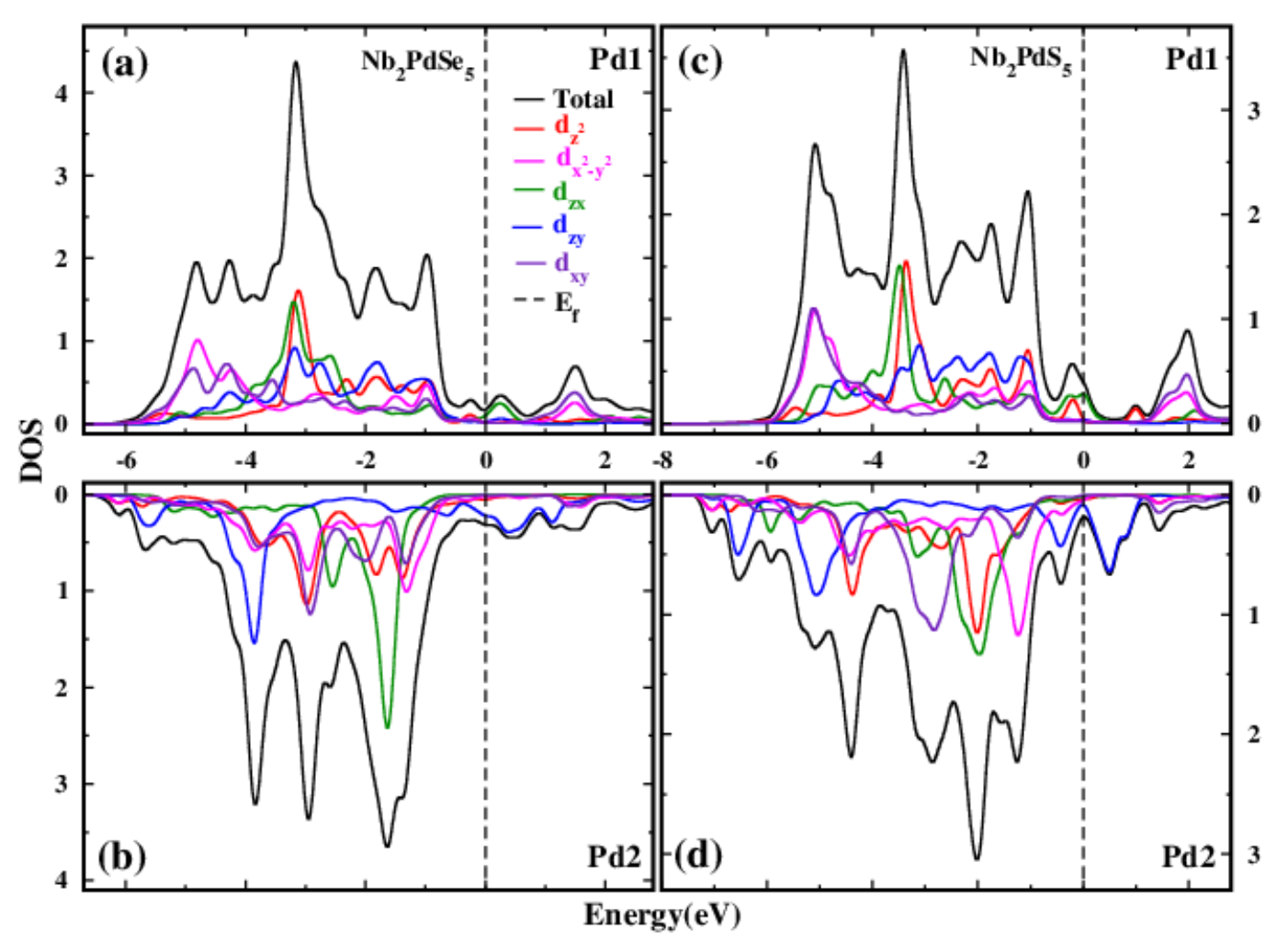}
\caption{\label{pddos}(a) and (b) DOS originated from different orbitals of Pd-4d at 
sites Pd1 and Pd2 respectively in Nb$_2$PdSe$_5$. (c) and (d) show the same
 DOS at Pd1 and Pd2 site respectively in Nb$_2$PdS$_5$.}
\end{figure}

As clear from the above results of  DOS  that the Pd-4d, and Nb-4d
states are strongly hybridized with Se/S states. The hybridization could be differ
at two distinct coordination sites Pd1 and Pd2 due to different structural environment.
In order to provide more insights into the effects of distinct 
coordination environment  m-DOS of Pd is presented in Fig.\ref{pddos}.
 Fig.\ref{pddos} (a) and (b) show Pd-4d states along with different d orbital contributions at Pd1 and Pd2 sites 
respectively in Nb$_2$PdSe$_5$. Fig.\ref{pddos} (c) and (d) show the same 
states in Nb$_2$PdS$_5$. The large band width of the Pd-4d seen in both the 
compounds indicates significant hybridization between the Pd-4d and the nn 
anion orbitals. In Nb$_2$PdSe$_5$, at the Pd1 site, the 4d$_{x^2-y^2}$ and 4d$_{xy}$ 
orbitals are directed towards the square planarly coordinated nn Se atoms and 
show largest separation between bonding (-4.81 eV and -4.28 eV) and 
antibonding (1.51 eV) states.  The 4d$_{3z^2-r^2}$ and 4d$_{zx}$ orbitals are oriented 
towards the nn Pd and Nb atoms and show comparatively smaller separation 
between bonding (-3.81 eV) and antibonding (0.27 eV) states. This indicates a 
weak hybridization possibly due to the larger inter-atomic distances Pd1-Pd1 
(3.406\AA{}) and Pd1-Nb1 (3.102\AA{}) in comparison to the Pd1-Se (2.392\AA{}). 
On the other hand, at the Pd2 site, square planar coordination of the nn Se 
atoms are aligned in the YZ plane (Fig.\ref{crystal}). Therefore, the 
4d$_{yz}$ orbital is strongly hybridized with the nn Se and antibonding 
states of this orbital are predominant near the E$_f$. States originate from 
4d$_{3z^2-r^2}$, 4d$_{xy}$ and 4d$_{x^2-y^2}$ orbitals are moderately 
hybridized while the 4d$_{zx}$ states show atomic like character. 
Apart from the differences in DOS of individual 4d orbitals originating from distinctly
coordinated Pd atoms(Pd1, and Pd2), the largest peak of Pd1-4d DOS lies at higher BE 
relative to same peaks in Pd2-4d DOS.  These differences could be related to different 
interorbital hybridization strengths due to the smaller distance of nn Se 
atoms at the Pd1 site (2.392 \AA{}) as compared to the Pd2 site (2.598\AA{}). 
In case of Nb$_2$PdS$_5$, the orbital character of different 4d orbitals of 
Pd at both the sites are similar to those of Nb$_2$PdSe$_5$.
However, the DOS structure of various 4d orbitals for both Pd1 and Pd2 atoms are 
slightly different from Nb$_2$PdSe$_5$. This difference could be associated to  change
in the hybridization strength due to variation in the bond lengths  between the two compounds.
These results are compared with experimental findings which informs that the higher BE
VB features G, and F(Fig.\ref{valcom}) fall in bonding states of Pd1(4d$_{x^2-y^2}$ and 4d$_{xy}$) 
while the 4d$_{3z^2-r^2}$ (-1.65 eV) and 4d$_{x^2-y^2}$ (-1.35 eV) states of Pd2 are 
close to the experimentally observed features C and B respectively 
(Fig.\ref{valcom}) in Nb$_2$Pd$_{1.2}$Se$_5$. Likewise, 
the energy position of differently coordinated Pd1 (4d$_{x^2-y^2}$ and 4d$_{xy}$) and Pd2 
(4d$_{zy}$) states at BE -5.12 eV and -2.09 eV matches closely with the experimental 
features G and D respectively in the VB spectra of Nb$_2$Pd$_{0.95}$S$_5$ 
(Fig.\ref{valcom}). Thus, the qualitative similarity of the experimental data with the calculated 
DOS could  establish the presence of  differently coordinated Pd atoms in both the compounds
which  has also been predicted theoretically in previous reports\cite{Zhang,Singh,khim}.

\begin{figure}
\includegraphics[width=10cm,keepaspectratio]{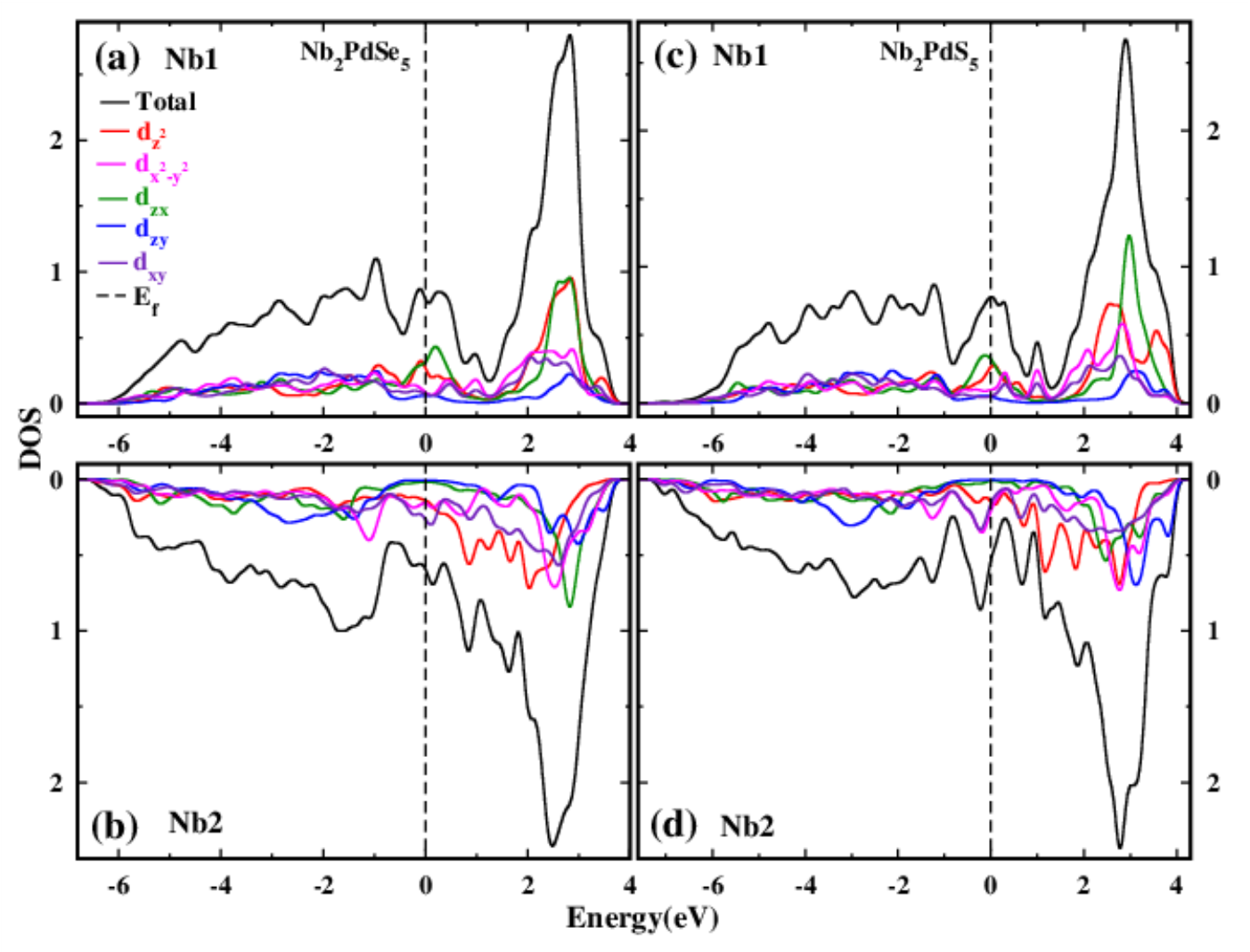}
\caption{\label{nbdos}DOS originated from different orbitals of Nb-4d at 
site Nb1 and Nb2 are shown in (a) and (b) respectively in Nb$_2$PdSe$_5$. 
(c) and (d) show same DOS at Nb1 and Nb2 site respectively in Nb$_2$PdS$_5$.}
\end{figure}

In both the compounds Nb atoms also possess distinct coordination environment
at two inequivalent sites Nb1 and Nb2, like Pd atoms. We  investigate the role   
of coordination geometry  in the DOS of Nb atoms. Fig.\ref{nbdos} depicts the 
Nb-4d DOS with different d orbital contributions at Nb1 and Nb2 sites for 
Nb$_2$PdSe$_5$(Fig.\ref{nbdos}(a) and (b)) and Nb$_2$PdS$_5$(Fig.\ref{nbdos}
(c) and (d)). In both the cases, the DOS of Nb1 and Nb2 is not appreciably differed, unlike the case of Pd DOS.
This could be the result of nearly same inter-atomic distances of the Nb atoms with their
nn Se/S atoms at both the coordination sites(Nb1 and Nb2).
However, a slight difference is visible  between the DOS structure of Nb1 and Nb2 in the vicinity of E$_f$ which
could be an important factor in the electronic structure of these compounds.
Further, the orbital character of states which are predominantly involved in the E$_f$ crossing is
4d$_{3z^2-r^2}$, and 4d$_{zx}$ orbitals of Nb1 and  4d$_{xy}$ of Nb2.
This result, as well as similar behaviour of the E$_f$ crossing by orbital specific
Pd-4d DOS(see Fig.\ref{pddos}) are signature of multiband nature of these compounds which is consistent with the observed 
multigap behavior in heat capacity measurements on these compounds\cite{Jha,khim}. 
These calculated results are also in good agreement with previous 
reports\cite{Singh,Zhang,khim}.

\begin{figure}
\includegraphics[width=10cm,keepaspectratio]{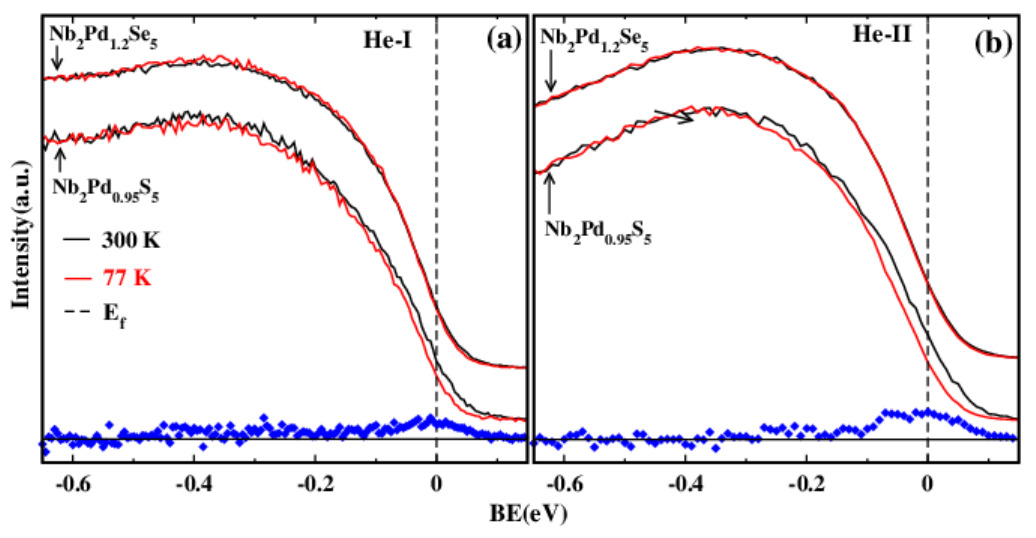}
\caption{\label{valfer}Comparison between 300K (Black) and 77K (Red) near 
E$_f$ VB spectra, of Nb$_2$Pd$_{1.2}$Se$_5$ and Nb$_2$Pd$_{0.95}$S$_5$ taken 
at (a) HeI and (b) HeII source energy. Blue dotted points represent
the difference between spectra  collected at 300K and 77K of Nb$_2$Pd$_{0.95}$S$_5$. }
\end{figure}

Fig.\ref{valfer}(a) and (b) show a comparison between the VB spectra of 
Nb$_2$Pd$_{1.2}$Se$_5$ and Nb$_2$Pd$_{0.95}$S$_5$ at 300 K and 77 K using He 
I and He II excitation energy, where differences between the normalized
 spectra collected at 300 K and 77 K are also plotted in each graph in order to 
highlight the temperature dependent changes. In case of Nb$_2$Pd$_{1.2}$Se$_5$, the near 
E$_f$ states remain unchanged with the lowering of temperature. On the other 
hand, Nb$_2$Pd$_{0.95}$S$_5$ spectra shows a depletion of spectral weight near 
the E$_f$ which is more pronounced in case of He II. This suppression of DOS 
is a signature of a pseudogap. In addition, an earlier study shows a crossover of transport carriers 
from electron to hole has been found at 100 K in Hall measurements in this compound\cite{Jha}.
These two observations are  consistent with  previous report describing  a pseudogap 
driven sign reversal in Hall coefficients\cite{foll}. On the other hand, in Nb$_2$Pd$_{1.2}$Se$_5$ 
the energy scale T $\sim$ 50 K, below which a crossover in the electronic 
ground state has been realized in the transport measurements\cite{khim} is 
lower than our measurement temperature 77 K. This could possibly be the 
reason for the unseen pseudogap feature at 77 K in Nb$_2$Pd$_{1.2}$Se$_5$.

\section{Conclusion}

We have presented a comparative study of the VB electronic structure 
of Nb$_2$Pd$_{1.2}$Se$_5$ and Nb$_2$Pd$_{0.95}$S$_5$ in conjugation with DFT based 
calculations. We find the VB spectra of both the compounds are qualitatively similar, though all
the features are observed at slightly higher BE in Nb$_2$Pd$_{0.95}$S$_5$ relative to Nb$_2$Pd$_{1.2}$Se$_5$
which could be the result of larger electronegativity of S anion in comparison to
Se anion. The calculated DOS describe the VB features are mainly composed 
of Pd-Se/S hybridized states. The different nature of DOS originating from the differently coordinated atoms
signifies an important role of the complex structural geometry in the electronic structure 
of these compounds. In addition, in our calculated DOS, states crossing the E$_f$ are dominated 
by different Pd-4d and Nb-4d orbitals ensuring significant role of multiband effects in these compounds.
 Furthermore, Nb$_2$Pd$_{0.95}$S$_5$ spectra exhibits a depletion of spectral weight near the E$_f$
with lowering the temperature to 77 K which results a pseudogap feature. This observation is consistent with previous finding
 of sign reversal in Hall coefficients below the temperature 100 K. On the other hand, temperature dependent pseudoagap is absent in 
case of Nb$_2$Pd$_{1.2}$Se$_5$ which could be due to  the energy scale T $\sim$ 50 K of crossover in the electronic 
ground state  is lower than our measurement temperature 77 K. 
Our comprehensive valence band electronic structure 
study in these compounds may open ways for further experimental and theoretical investigations in this 
field.


\end{document}